# Proposal submitted to Danish Science Foundation Fall 2017

## Guiding Chemical Synthesis:
## Computational Prediction of the Regioselectivity of CH functionalization

Jan H. Jensen, University of Copenhagen

## 1. Introduction: The importance of predicting regioselectivity of C-H functionalization

The synthesis of new organic molecules is central to the field of chemistry with important applications such as materials design and drug discovery. But synthesis is usually a time-consuming and, hence, expensive process and increasing the efficiency of chemical synthesis is a grand challenge within chemistry.[1] CH functionalization i.e. replacing the hydrogen atom in a CH bond with another atom or molecule, is arguably single most promising technique for increasing the efficiency of chemical synthesis. According to one recent editorial[2] (emphasis added) "CH functionalization is revolutionizing the way that organic chemists approach the synthesis of target molecules. … CH functionalization represents a paradigm shift from the standard logic of organic synthesis. Instead of focusing on orchestration of selective reactions at functional groups, the new logic relies on the **controlled** functionalization of specific CH bonds, even in the presence of supposedly more reactive functional groups". "Controlled" is the operative word here: CH functionalization is only useful as synthetic tool when the reaction happens at one site (i.e. it is regioselective) that can be reliably predicted. However, as discussed next, there is no general predictive tool for this important problem. We propose to develop such a predictive tool and demonstrate that it can be used to make synthetic strategies more efficient. To meet this goal several fundamental questions regarding chemical reactivity of CH bonds must be answered, which makes FNU the most appropriate council to consider this proposal.

      Prof. Jan Jensen (primary PhD supervisor) and Prof. Jesper Kristensen and a PhD student (to be hired) will carry out the theoretical and experimental studies at the Department of Chemistry and Department of Drug Design and Pharmacology at KU, respectively.

## 2. State-of-art in predicting regioselectivity of C-H functionalization

The prediction of the regioselectivity of CH functionalization is usually done using heuristic rules that can be found in most organic chemistry textbooks, such as the well-known 3° > 2° > 1° stability order for $sp^3$ radical centers. However, for more complicated molecules with many similar CH sites this approach becomes difficult to apply with confidence. A few researchers have developed computational methods but they are either heavily parameterized and applicable only to specific catalysts[3] or catalysts that are not widely used.[4] One exception is the recent work by Jørgensen and co-workers[5] that predicts the regioselectivity of electrophilic aromatic substitution reactions of heteroaromatic systems, but unfortunately, this approach fails for several heteroaromatic systems of interest. There are, of course, a myriad of density functional theory (DFT) studies on CH functionalization[6] but such studies typically look the reaction only at the sites identified experimentally and are rarely, if ever, used to predict regioselectivity to *guide* synthesis. As a result, synthetic chemists usually predict regioselectivity using textbook rules-of-thumb that often fail for complex molecules. There is thus a great need for efficient and reliable computational tools that can predict regioselectivity for complex molecules without prior extensive parameterization for a particular catalyst and this is what we propose to develop.

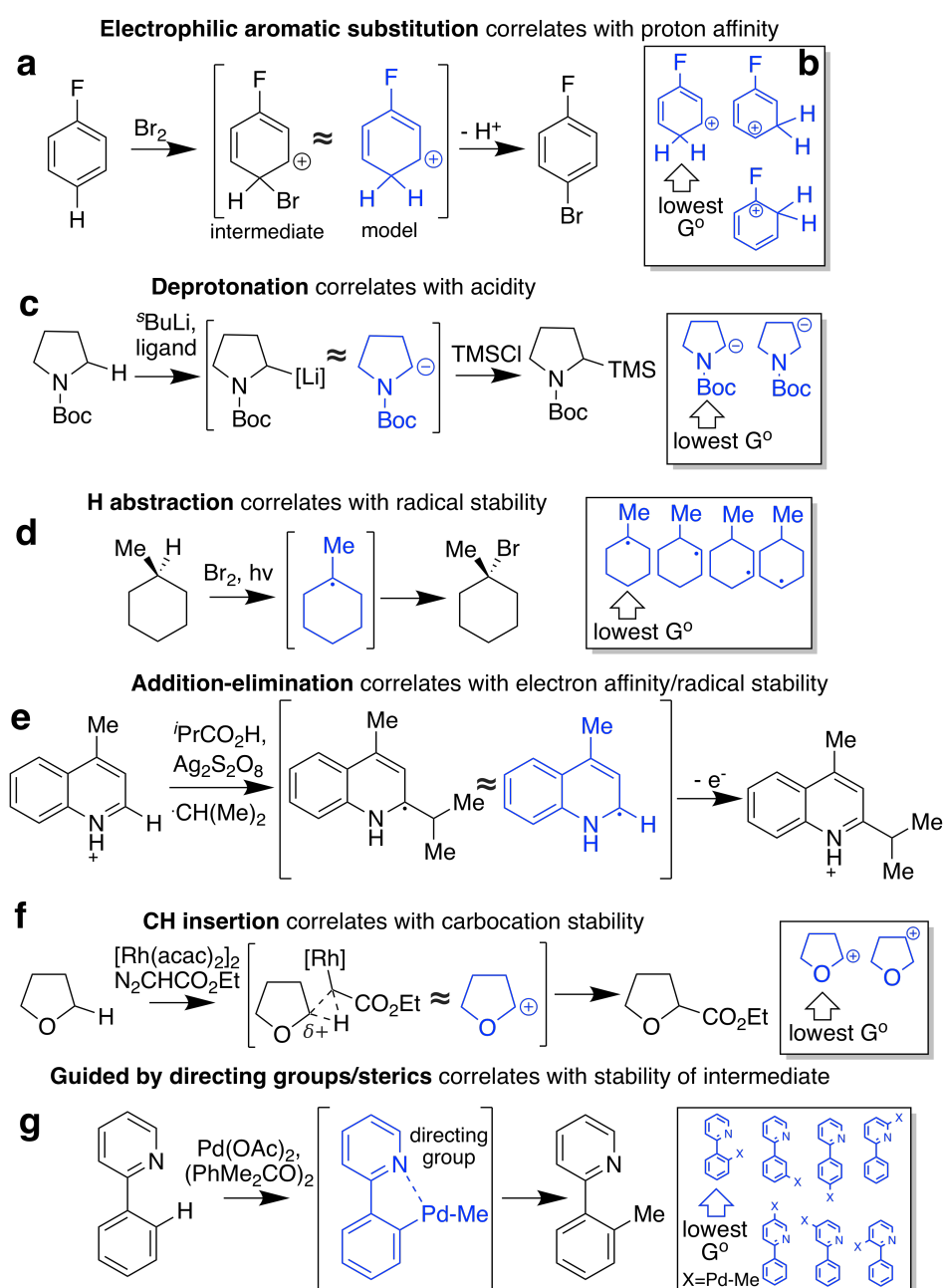

**Figure 1**. The most common mechanisms of CH functionalization and their correlation with easily computed molecular properties of intermediates or models thereof (in blue). Most examples taken from ref [7]. **Preliminary calculations are able to reproduce the observed regioselectivities**. The current version of regiosqm.org can only predict proton affinities (Figure 1a-b)

# 3. The project
## 3.1. Objectives & Methodology.

**Objective 1.** Develop a **general** and user-friendly method to predict regioselectivity

**Objective 2**. Demonstrate that our method can make chemical synthesis more efficient

**Preliminary Results.** The basic idea behind our proposed work is illustrated in Figure 1a-b. The hypothesis is that electrophilic aromatic substitution (EAS) occurs at the CH center with the highest proton affinity[8] (Figure 1a) and that the regioselectivity of EAS reactions can be predicted by creating all protonated isomers and finding the one with the lowest free energy (Figure 1b). The free energies using semiempirical quantum mechanics, which is ca 1000 times faster than conventional DFT, and each prediction requires only 2-20 minutes of CPU time depending on the size of the molecule. We tested this approach (which we call RegioSQM) on over 500 reactions with a 95% success rate[9] and made it available as an easy-to-use web server at [regiosqm.org](regiosqm.org). W**e propose to use a similar approach for the remaining main types of CH functionalization reactions** by discovering which molecular property is the primary determinant of reactivity for each reaction type (Figure 1).

To demonstrate the utility of RegioSQM in chemical synthesis we will show that it can be used in **late stage functionalization** (LSF) of potential new drugs (WP4). In the late synthetic stages of drug discovery, a promising molecule has been found whose properties must be optimized by adding different functional group to a one or two specific CH groups out of dozens possible. RegioSQM can predict which CH groups will be functionalized under certain conditions, thereby obviating trial and error on very costly starting material.

**Workpackages:** Jensen is WP leader for WP1-3 and Kristensen for WP4.

**WP1. Extension of RegioSQM to predict regioselectivity of deprotonation and H abstraction reactions** (Figure 1c-d). We will assemble a test (roughly 100 examples) and validation set (roughly 300 examples) of experimentally observed regioselectivity for data for each type of reaction the chemical literature (**T1.1**). The test set will be used to find the best computational model for each type of reaction. Following our previous work on $pK_a$ prediction,[10] the H affinity and acidity will be computed relative to experimental reference data for small representative molecules. The optimal composition of this reference set will be determined using the training data set (**T1.2**). **Deliverable**: RegioSQM v1.1 release and submission of paper (**D1.1**). **Risk assessment**: **low**. The correlation between regioselectivity for these reactions and radical stability and acidity is well established as are the ability of QM methods to predict these quantities reliably.[11]

**WP2. Extension of RegioSQM to predict regioselectivity of addition-elimination and CH insertion** (Figure 1e-f). The tasks are similar to the corresponding tasks for WP1 (**T2.1** and **T2.2**). **Deliverable**: RegioSQM v1.2 release and submission of paper (**D2.1**). In the case of electron affinity (Figure 1e), the position of the radical will be determined by a Mulliken analysis of the spin-density. **Risk assessment**: **low/medium**. The correlation between regioselectivity for these reactions and carbocation stability and electron affinity is less well established, although our preliminary results are encouraging (Figure 1), and **discovering the chemical model/property that governs regioselectivity for each of these reactions is a fundamental scientific challenge**. Should the approach fail, the next step will be to determine the actual barrier heights in an automated fashion.[12] SQM methods are generally less reliable for barrier heights, so it is possible we will have to use more accurate, but more computationally intensive methods, such as DFT. In this case, it will take hours instead of minutes to reach the accuracy that RegioSQM has for EAS reactions, but the method is still of practical use in planning synthetic strategies.

**WP3. Extension of RegioSQM to predict regioselectivity of guided reactions** (Figure 1g). The tasks are similar to the corresponding tasks for WP1 (**T2.1** and **T2.2**). Since we are modeling reaction intermediates for which there is little experimental data, we will build a library of reference values using high level quantum mechanical calculations. **Deliverable**: RegioSQM v2.0 release and submission of paper (**D3.1**). **Risk assessment**: **medium/high**. Guided regioselectivity will be catalyst specific to some degree and **whether it is possible to find representative catalyst models that work in general is a fundamental scientific question**. Just as for WP2 we may end up computing barrier heights instead with computationally more demanding methods.

**WP4. Use of RegioSQM to functionalize complex molecules.** We will select 2-3 drug-like molecules that each contain several types of CH bonds, use RegioSQM to predict the regioselectivity using different reaction types and then carry out the reactions to verify the predictions (**T4.1**). We will also determine how many different CH bonds can be selectively functionalized by, for example, adding protecting groups to the most reactive CH sites of a particular type to target the less reactive CH sites of the same type (**T4.2**). **Deliverable**: Submission of paper (**D4.1**). **Risk assessment: low/medium** (T4.1/T4.2)**.** Since the goal is to verify computational predictions rather than to make a particular molecule we can focus on molecules that are found easy to work with. We will select commercially available molecules (e.g. quinine and codeine) to allow for quick and extensive testing.

**3.2 Expected outcomes & impacts and originality** The expected outcome will be the first computational tool (RegioSQM) that can predict which CH bond in a molecule is most reactive under different conditions. RegioSQM is freely available and can be used by synthetic chemists who are not experts in computational chemistry, through at a user-friendly web interface at regiosqm.org, to greatly increase the efficiency of chemical synthesis. This will ultimately lower the R&D cost of, and may lead to new discoveries in, drug discovery and materials design.

## 4. Feasibility of the project
### 4.1. Work plan

|  | 2018 |  | 2019 |  | 2020 |  |  | 2021 |  |
|---|---|---|---|---|---|---|---|---|---|
| Month | 4 | 8 | 12 | 16 | 20 | 24 | 28 | 32 | 36 |
| **WP1** | T1.1 | T2.1,D1.1 |  |  |  |  |  |  |  |
| **WP2** |  |  | T2.1 | T2.2,D2.1 |  |  |  |  |  |
| **WP3** |  |  |  |  | T3.1 | T3.2 | T3.2 | D3.1 |  |
| **WP4** |  |  |  |  | T4.1 | T4.1&2 | T4.1&2 | T4.1&2 | D4.1 |

**4.2. Competences and infrastructure** Jensen has more than 20 years of experience in the development of computational models and web servers as outlined in his CV. He is deputy head of department in charge of teaching until ultimo 2018 after which he can devote the majority of his time to the proposed work. He has written most of the RegioSQM code that the project will build on. Jensen has access to 1500 cores as part of the University of Copenhagen branch of the Danish Supercomputer Center. The DCSC provides a full-time TAP for the day-to-day running of the DCSC/KU center. The semiempirical calculations require relatively modest computation resources so this is more than adequate.

The Kristensen-lab has been involved with projects dealing with the selective functionalization of a numerous of different scaffolds within various medicinal chemistry projects at the Department of Drug Design and Pharmacology at the University of Copenhagen. All the necessary infrastructure and knowhow to conduct the experimental work is available: state-of-the-art laboratories with extensive analytical capabilities (LC-MS-MS and NMR) allowing for quick quantification and structure elucidation.

A PhD student will be hired at the start of the grant and will need a background in computational chemistry and programming with some experience in synthetic chemistry desirable and we already have several candidates in mind so recruitment will not be a problem.

**4.3. Budget** The salary and taxameter for the PhD student represents most of the budget. 30K DKK are requested to buy a dedicated computer for the web-server.

**4.4. Ethics** The proposed work will not involve the use of laboratory animals.

**5. Dissemination and communication of results** All theoretical developments and applications will be published in open access peer-reviewed journals (e.g. *Chemical Science, Beilstein Journal of Organic Chemistry*) to allow access to as many people as possible. All software will be made freely available to the scientific community as a user-friendly software package under a permissive license and as a web interface at regiosqm.org. The PhD students will co-supervise at least 5 MSc students thereby transferring knowledge. Finally, we will present our results at a Gordon Research Conference on Organic Reactions and Processes and at an Internal Symposium on CH Activation, both held annually.

**References**

[1] Young, I. S., & Baran, P. S. (2009). Protecting-group-free synthesis as an opportunity for invention. *Nature Chemistry*, *1*(3), 193-205.

[2] Davies, H. M., & Morton, D. (2016). Recent advances in C–H functionalization. *Journal of Organic Chemistry* 81, 343−350

[3] (a) Gormisky, P. E., & White, M. C. (2013). Catalyst-controlled aliphatic C–H oxidations with a predictive model for site-selectivity. *Journal of the American Chemical Society*, *135*(38), 14052-14055. (b) Bess, E. N., Guptill, D. M., Davies, H. M., & Sigman, M. S. (2015). Using IR vibrations to quantitatively describe and predict site-selectivity in multivariate Rh-catalyzed C–H functionalization. *Chemical Science*, *6*(5), 3057-3062.

[4] Margrey, K. A., McManus, J. B., Bonazzi, S., Zecri, F., & Nicewicz, D. A. (2017). Predictive Model for Site-Selective Aryl and Heteroaryl C–H Functionalization via Organic Photoredox Catalysis. *Journal of the American Chemical Society*, *139*(32), 11288-11299.

[5] Kruszyk, M., Jessing, M., Kristensen, J. L., & Jørgensen, M. (2016). Computational Methods to Predict the Regioselectivity of Electrophilic Aromatic Substitution Reactions of Heteroaromatic Systems. *The Journal of organic chemistry*, *81*(12), 5128-5134.

[6] (a) Balcells, D., Clot, E., & Eisenstein, O. (2010). C—H Bond Activation in Transition Metal Species from a Computational Perspective. *Chem. Rev*, *110*(2), 749-823. (b) Musaev, D. G., Figg, T. M., & Kaledin, A. L. (2014). Versatile reactivity of Pd-catalysts: mechanistic features of the mono-N-protected amino acid ligand and cesium-halide base in Pd-catalyzed C–H bond functionalization. *Chemical Society Reviews*, *43*(14), 5009-5031.

[7] Cernak, T., Dykstra, K. D., Tyagarajan, S., Vachal, P., & Krska, S. W. (2016). The medicinal chemist's toolbox for late stage functionalization of drug-like molecules. *Chemical Society Reviews*, *45*(3), 546-576.

[8] (a) Streitwieser, A. *Molecular Orbital Theory for Organic Chemists*, John Wiley & Sons Inc, 1961. (b) Wang, D. Z., & Streitwieser, A. (1999). An ab initio study of electrophilic aromatic substitution. *Theoretical Chemistry Accounts: Theory, Computation, and Modeling (Theoretica Chimica Acta)*, *102*(1), 78-86.

[9] Kromann, J. C., Jensen, J. H., Kruszyk, M., Jessing, M. & Jørgensen, M Fast and Accurate Prediction of the Regioselectivity of Electrophilic Aromatic Substitution Reactions. *Chemical Science*, submitted. Preprint: https://doi.org/10.26434/chemrxiv.5435935.v1

[10] Jensen, J. H., Swain, C. J., & Olsen, L. (2017). Prediction of p K a Values for Druglike Molecules Using Semiempirical Quantum Chemical Methods. *The Journal of Physical Chemistry A*, *121*(3), 699-707.

[11] Shen, K., Fu, Y., Li, J. N., Liu, L., & Guo, Q. X. (2007). What are the pKa values of C–H bonds in aromatic heterocyclic compounds in DMSO? *Tetrahedron*, *63*(7), 1568-1576.

[12] Zimmerman, P. M. (2015). Navigating molecular space for reaction mechanisms: an efficient, automated procedure. *Molecular Simulation*, *41*(1-3), 43-54.